\documentclass[11pt,a4]{article}

\usepackage{bm,amsmath,amssymb,a4wide,wasysym,amsfonts,cite,esint}

\usepackage[hyperindex,breaklinks]{hyperref}

\begin{document}

\title{\textbf{Equilibrium electric current of massive electrons with anomalous magnetic moments induced by a magnetic field and the electroweak interaction with matter}}

\author{Maxim Dvornikov$^{a,b}$\thanks{maxdvo@izmiran.ru}
\\
$^{a}$\small{\ Pushkov Institute of Terrestrial Magnetism, Ionosphere} \\
\small{and Radiowave Propagation (IZMIRAN),} \\
\small{108840 Troitsk, Moscow, Russia;} \\
$^{b}$\small{\ Physics Faculty, National Research Tomsk State University,} \\
\small{36 Lenin Avenue, 634050 Tomsk, Russia}}

\date{}

\maketitle

\begin{abstract}
We study the possibility of the existence of the electric current,
formed by massive electrons and positrons, flowing along an external magnetic
field. The charged fermions are supposed to have nonzero anomalous
magnetic moments and electroweakly interact with background matter.
The expression for the current is obtained on the basis of the exact
solution of the Dirac equation in the corresponding external fields.
We demonstrate that, in the state of equilibrium, such a current is vanishing for any characteristics
of the electron-positron plasma as well as the external fields. Our
results are compared with the recent findings of other authors.
\end{abstract}

\section{Introduction\label{sec:INTR}}

The dynamo mechanism is widely used in cosmology and astrophysics
for the generation of strong large-scale magnetic fields~\cite{ZelRuzSok83}.
This mechanism is based on the excitation of the electric current
along the external magnetic field $\mathbf{B}$: $\mathbf{J}=\Pi\mathbf{B}$. Indeed,
if a current $\mathbf{J}\parallel\mathbf{B}$ is accounted for in the Maxwell
equations, the magnetic field becomes unstable and can be dynamo amplified.
In QED plasma, where the parity is conserved, the parameter $\Pi$ should be a pseudoscalar since $\mathbf{J}$ is a vector and $\mathbf{B}$ is an axial-vector.
For example, in classical MHD, the parameter $\Pi\sim\left\langle \mathbf{v}\cdot(\nabla\times\mathbf{v})\right\rangle $~\cite{Sti04},
where $\mathbf{v}$ is the random component of the fluid velocity, $\langle \mathbf{v} \rangle = 0$. Thus, one can see that the situation when $\Pi\neq0$ is implemented effectively in a medium.

Recently, the dynamo mechanism based on the chiral magnetic effect
(the CME)~\cite{Vil80,NieNin81} becomes popular. The CME consists
in the appearance of the current $\mathbf{J}_{\mathrm{CME}}=\alpha_{\mathrm{em}}(\mu_{\mathrm{R}}-\mu_{\mathrm{L}})\mathbf{B}/\pi$
of massless charged particles, as the consequence of the Adler-Bell-Jackiw anomaly in QED~\cite{FukKhaWar08}. Here $\mu_{\mathrm{R,L}}$ are the chemical
potentials of right and left chiral fermions and $\alpha_{\mathrm{em}}\approx1/137$
is the fine structure constant. We remind that the chiral imbalance $\mu_5=(\mu_{\mathrm{R}} - \mu_{\mathrm{L}})/2$ is the pseudoscalar under the spatial inversion, $\mu_{\mathrm{R}} \leftrightarrow \mu_{\mathrm{L}}$, i.e. the parity is conserved in QED for $\mathbf{J}_{\mathrm{CME}}$. The manifestations of the CME are
extensively studied in astrophysics and cosmology~\cite{Sig17}, accelerator experiments~\cite{Kha15}, and solid state physics~\cite{ArmMelVis}.
The correction to the CME from the parity violating electroweak interaction was considered for the first time in Ref.~\cite{BoyRucSha12}.

The main feature of the CME is the unbroken chiral symmetry of charged
particles. It means that any nonzero mass makes $\mathbf{J}_{\mathrm{CME}}$
to vanish~\cite{Vil80}. The majority of known elementary particles
acquire masses through the electroweak mechanism. However, it is likely
to be an electroweak crossover rather than a first order phase transition~\cite{LaiMey15}.
Therefore, charged particles will remain massive at any pressure and
chemical potential unless a new physics beyond the standard model
is accounted for (see, e.g., Ref.~\cite{CliKaiTuc17}). There are indications that a chiral phase transition
can happen in dense matter owing to the QCD effects~\cite{Smi97}.
Some astrophysical applications for the magnetic fields generation
in compact stars due to the CME and the electroweak interaction between
quarks in dense matter are considered in Ref.~\cite{Dvo17b}.

In this connection, there is a particular interest in searching for
the possibility of the generation of the current $\mathbf{J}\parallel\mathbf{B}$
in the system of massive charged particles. In this situation, one
would have an instability of the magnetic field without necessity
to demand the restoration of the chiral symmetry. One of the examples
of such a system was studied in Ref.~\cite{SemSok04}, where the
dynamo amplification of magnetic fields in an inhomogeneous electroweak matter
was discussed.

Recently, in Refs.~\cite{BubGubZhu17,Dvo17a}, the existence of an
electric current $\mathbf{J}\parallel\mathbf{B}$ of massive fermions,
having anomalous magnetic moments and interacting with an external axial-vector
field, was claimed. The axial-vector field can be represented
in the form of the electroweak background matter~\cite{Dvo17a} or
in a hypothetical CPT-odd extension of the standard model~\cite{BubGubZhu17}.
In both cases, a nonzero $\mathbf{J}\parallel\mathbf{B}$ was shown to
exist in the considered system.

In the present work, we revise the results of Refs.~\cite{BubGubZhu17,Dvo17a}.
In Sec.~\ref{sec:CANCCUR}, we demonstrate that the current $\mathbf{J}\parallel\mathbf{B}$
of massive particles with anomalous magnetic moments, interacting
with the parity violating axial-vector field, is vanishing. This result
is based on the computation of this current using the exact solution
of the Dirac equation in the corresponding external fields performed
in Appendix~\ref{sec:POSCURR} (see also Refs.~\cite{BalStuTok12,BalStuTok13}).
Our results are summarized in Sec.~\ref{sec:CONCL}.

\section{Cancellation of the anomalous electric current\label{sec:CANCCUR}}

Let us consider a plasma of electrons and positrons, electroweakly
interacting with background matter under the influence of the external
magnetic field $\mathbf{B}$. Charged particles are considered to
be massive. Their nonzero anomalous magnetic moments are taken into
account. The Lagrangian for an electron, described by the bispinor
$\psi_{e}$, has the form,
\begin{equation}\label{eq:Larg}
  \mathcal{L} = \bar{\psi}_{e}
  \left[
    \gamma_{\mu}(\mathrm{i}\partial^{\mu}+eA^{\mu}) - m +
    \frac{\mu}{2}\sigma^{\mu\nu}F_{\mu\nu} -
    \gamma_{\mu}(V_{\mathrm{L}}^{\mu}P_{\mathrm{L}} +
    V_{\mathrm{R}}^{\mu}P_{\mathrm{R}})
  \right]
  \psi_{e},
\end{equation}
where $A^{\mu}=(0,0,Bx,0)$ is the vector potential corresponding
to the constant and homogeneous magnetic field, directed along the
$z$-axis, $e>0$ is the elementary charge, $P_{\mathrm{L,R}}=(1\mp\gamma^{5})/2$
are the chiral projectors, $\gamma^{\mu}=(\gamma^{0},\bm{\gamma})$,
$\sigma_{\mu\nu}=\tfrac{\mathrm{i}}{2}[\gamma_{\mu},\gamma_{\nu}]_{-}$,
and $\gamma^{5}=\mathrm{i}\gamma^{0}\gamma^{1}\gamma^{2}\gamma^{3}$
are the Dirac matrices, $m$ is the electron mass, $\mu$ is the anomalous
magnetic moment~\cite{Odo06}, $F_{\mu\nu}=\partial_{\mu}A_{\nu}-\partial_{\nu}A_{\mu}=(\mathbf{E},\mathbf{B})$
is the electromagnetic field tensor (with $\mathbf{E}=0$), and $V_{\mathrm{L,R}}^{\mu}=(V_{\mathrm{L,R}}^{0},\mathbf{V}_{\mathrm{L,R}})$
are the effective potentials of the electroweak interaction of the
electron chiral projections with background matter. We shall suppose
that the background matter is macroscopically at rest and unpolarized.
In this situation, $\mathbf{V}_{\mathrm{L,R}}=0$ and $V_{\mathrm{L,R}}^{0}\equiv V_{\mathrm{L,R}}\neq0$.
The explicit form of $V_{\mathrm{L,R}}$ is given in Ref.~\cite{DvoSem15a}
for the case of background matter consisting of neutrons and protons.

The most general expression for the induced current of electrons $\mathbf{J}_{e}$
was obtained in Ref.~\cite{Dvo17a} (see also Eq.~(\ref{eq:Jpn>0})).
The electric current of positrons $\mathbf{J}_{\bar{e}}$ can be derived
on the basis of the exact solution of the Dirac equation for positrons
$\psi_{\bar{e}}$, accounting for the external fields, which, in its
turn, can be obtained from $\psi_{e}$ by applying the charge conjugation.
The details of the derivation of $\mathbf{J}_{\bar{e}}$ are provided
in Appendix~\ref{sec:POSCURR}.

Summing up the induced currents of electrons (see Eq.~(\ref{eq:Jpn>0})
and Ref.~\cite{Dvo17a}), and positrons (see Eqs.~(\ref{eq:Jposn>0})),
one gets the following expression for the total current $\mathbf{J}=\mathbf{J}_{e}+\mathbf{J}_{\bar{e}}$:
\begin{equation}\label{eq:Jtot}
  \mathbf{J}=\Pi\mathbf{B},
  \quad
  \Pi=-\frac{\alpha_{\mathrm{em}}}{\pi}
  \sum_{n=1}^{\infty}
  \sum_{s=\pm1}
  \int_{-\infty}^{+\infty}
  \frac{\mathrm{d}p_{z}}{\mathcal{E}}
  \left[
    p_{z}
    \left(
      1+s\frac{V_{5}^{2}}{R^{2}}
    \right) -
    s\frac{\mu BmV_{5}}{R^{2}}
  \right]
  \Delta f,
\end{equation}
where $\alpha_{\mathrm{em}}=e^{2}/4\pi$ is the fine structure constant,
$V_{5}=(V_{\mathrm{L}}-V_{\mathrm{R}})/2$, $\Delta f=f(\mathcal{E}-\chi_{\mathrm{eff}})-f(\mathcal{E}+\chi_{\mathrm{eff}})$,
$f(E)=\left[\exp(\beta E)+1\right]^{-1}$ is the Fermi-Dirac distribution
function, $\chi_{\mathrm{eff}}=\chi-\bar{V}$, $\chi$ is the chemical
potential of the electron-positron plasma, $\beta=1/T$ is the reciprocal
temperature, $\bar{V}=(V_{\mathrm{L}}+V_{\mathrm{R}})/2$, $n=1,2,\dotsc$
and $s=\pm1$ are discrete quantum numbers, which the energy levels
depend on (see Eqs.~(\ref{eq:En>0}) and~(\ref{eq:S2})), and
\begin{align}\label{eq:ER2}
  \mathcal{E} & =\sqrt{p_{z}^{2}+2eBn+m^{2}+(\mu B)^{2}+V_{5}^{2}+2sR^{2}},  
  \nonumber
  \\
  R^{2} & =\sqrt{(p_{z}V_{5}-\mu Bm)^{2}+2eBn
  \left[
    (\mu B)^{2}+V_{5}^{2}
  \right]}.
\end{align}
It is interesting to mention that the integrand in Eq.~(\ref{eq:Jtot})
can be represented as
\begin{equation}
  \frac{1}{\mathcal{E}}
  \left[
    p_{z}
    \left(
      1+s\frac{V_{5}^{2}}{R^{2}}
    \right) -
    s\frac{\mu BmV_{5}}{R^{2}}
  \right] =
  \frac{\partial\mathcal{E}}{\partial p_{z}}.
\end{equation}
Hence the total current is proportional to the averaged group velocity
of a charged particle along the magnetic field, $v_{z}=\partial\mathcal{E}/\partial p_{z}$:
$\mathbf{J}=-\alpha_{\mathrm{em}}\left\langle v_{z}\right\rangle \mathbf{B}/\pi$.

Considering Eq.~(\ref{eq:Jtot}) in case of a degenerate electron
gas, it was claimed in Ref.~\cite{Dvo17a} that $\Pi\neq0$. Analogous
result was obtained in Ref.~\cite{BubGubZhu17} on the basis of the
analysis of the effective Lagrangians in the one-loop approximation. The instability of
the magnetic field, driven by the anomalous current $\mathbf{J}=\Pi\mathbf{B}$,
and some astrophysical applications were studied in Ref.~\cite{Dvo17a}.
The claim of Refs.~\cite{Dvo17a,BubGubZhu17} that there is $\mathbf{J}=\Pi\mathbf{B}\neq0$
in the considered system is based on the fact that the energy levels
at $n>0$ in Eq.~(\ref{eq:ER2}) are neither symmetric nor antisymmetric
functions of $p_{z}$, which is the momentum projection along the magnetic field.
Thus, the integration over $p_{z}$ in the symmetric limits in Eq.~(\ref{eq:Jtot})
could give a nonzero result. The asymmetry coefficient of the energy
levels in Eq.~(\ref{eq:ER2}) is $\mu BmV_{5}$. It is this term,
which $\mathbf{J}\parallel\mathbf{B}$ in Refs.~\cite{Dvo17a,BubGubZhu17}
is proportional to.

Nevertheless, a careful analysis reveals that $\Pi=0$ in Eq.~(\ref{eq:Jtot}).
This fact is not quite obvious. To demonstrate it, we introduce the
notation in Eq.~(\ref{eq:Jtot}),
\begin{equation}\label{eq:Fdef}
  \frac{\Delta f}{\mathcal{E}}=F(Q^{2}+2sR^{2}),
  \quad
  Q^{2}=p_{z}^{2}+2eBn+m^{2}+(\mu B)^{2}+V_{5}^{2}.
\end{equation}
Then we decompose $F(Q^{2}+2sR^{2})$,
\begin{align}\label{eq:Fk}
  F(Q^{2}+2sR^{2}) & =
  \sum_{k=0}^{\infty}
  2^{2k}R^{4k}
  \left[
    \frac{F^{(2k)}(Q^{2})}{(2k)!}+2sR^{2}\frac{F^{(2k+1)}(Q^{2})}{(2k+1)!}
  \right],
  \nonumber
  \\
  F^{(k)}(Q^{2}) & =\frac{\mathrm{d}^{k}F(Q^{2})}{\mathrm{d}(Q^{2})^{k}},
\end{align}
in a formal series.

The sum over $s$ of the integrand in Eq.~(\ref{eq:Jtot}) gives
\begin{align}\label{eq:I1}
  I = & \sum_{s=\pm1}
  \left[
    p_{z}+s(p_{z}V_{5}-\mu Bm)\frac{V_{5}}{R^{2}}
  \right]
  F(Q^{2}+2sR^{2})
  \nonumber
  \\
  & = 2\sum_{k=0}^{\infty}2^{2k}R^{4k}
  \left[
    p_{z}\frac{F^{(2k)}(Q^{2})}{(2k)!}+2V_{5}(p_{z}V_{5}-\mu Bm)
    \frac{F^{(2k+1)}(Q^{2})}{(2k+1)!}
  \right]
  \nonumber
  \\
  & =2\sum_{k=0}^{\infty}2^{2k}
  \left\{
    p_{z}R^{4k}\frac{F^{(2k)}(Q^{2})}{(2k)!}+
    \frac{1}{k+1}\frac{\partial}{\partial p_{z}}
    \left[
      R^{4(k+1)}
    \right]
    \frac{F^{(2k+1)}(Q^{2})}{(2k+1)!}
  \right\},
\end{align}
where we use the identities,
\begin{equation}
  \frac{\partial R^{4}}{\partial p_{z}}=2V_{5}(p_{z}V_{5}-\mu Bm),
  \quad
  \frac{1}{k+1}\frac{\partial}{\partial p_{z}}
  \left[
    R^{4(k+1)}
  \right] =
  R^{4k}\frac{\partial R^{4}}{\partial p_{z}},
\end{equation}
and take into account Eq.~(\ref{eq:ER2}).

Integrating Eq.~(\ref{eq:I1}) over $p_{z}$ and then by parts, one
gets
\begin{align}\label{eq:I2}
  \int_{-\infty}^{+\infty}I\mathrm{d}p_{z} = &
  2\sum_{k=0}^{\infty}2^{2k}
  \Bigg[
    \int_{-\infty}^{+\infty}\mathrm{d}p_{z}
    \left\{
      R^{4k}p_{z}\frac{F^{(2k)}(Q^{2})}{(2k)!} -
      \frac{1}{k+1}\frac{\partial}{\partial p_{z}}
      \left[
        F^{(2k+1)}(Q^{2})
      \right]
      \frac{R^{4(k+1)}}{(2k+1)!}
    \right\}
    \nonumber
    \\
    & +
    \left.
      \frac{F^{(2k+1)}(Q^{2})R^{4(k+1)}}{(k+1)(2k+1)!}
    \right|_{-\infty}^{+\infty}
  \Bigg].
\end{align}
The function $F(Q^{2})$ is proportional to the Fermi-Dirac distribution
functions, which are vanishing at great values of the argument. The same
property has any derivative of $F(Q^{2})$ in Eq.~(\ref{eq:Fk}).
Thus, the last term in Eq.~(\ref{eq:I2}) disappears at $p_{z}\to\pm\infty$.

Taking into account that
\begin{equation}
  \frac{\partial}{\partial p_{z}}
  \left[
    F^{(2k+1)}(Q^{2})
  \right] =
  2p_{z}\frac{\mathrm{d}}{\mathrm{d}Q^{2}}
  \left[
    F^{(2k+1)}(Q^{2})
  \right] =
  2p_{z}F^{(2k+2)}(Q^{2}),
\end{equation}
and changing the summation index $k\to k-1$ in the second term in
Eq.~(\ref{eq:I2}), one obtains
\begin{align}\label{eq:I3}
  \int_{-\infty}^{+\infty}I\mathrm{d}p_{z}= & 
  2\sum_{k=0}^{\infty}2^{2k}
  \int_{-\infty}^{+\infty}\mathrm{d}p_{z}
  R^{4k}p_{z}\frac{F^{(2k)}(Q^{2})}{(2k)!} -
  2\sum_{k=1}^{\infty}2^{2k}
  \int_{-\infty}^{+\infty}\mathrm{d}p_{z}
  R^{4k}p_{z}\frac{F^{(2k)}(Q^{2})}{(2k)!}
  \nonumber
  \\
  & =
  2\int_{-\infty}^{+\infty}\mathrm{d}p_{z} \, p_{z}F(Q^{2})=0,
\end{align}
where we use the fact that $F^{(0)}(Q^{2})=F(Q^{2})$. To get the
vanishing result of the integration in the symmetric limits in Eq.~(\ref{eq:I3}),
we account for that $Q^{2}$ is the even function of $p_{z}$ (see
Eq.~(\ref{eq:Fdef})), i.e. the integrand in Eq.~(\ref{eq:I3})
is the odd function.

Thus, we have demonstrated that in Eq.~(\ref{eq:Jtot})
\begin{equation}\label{eq:canccur}
  \mathbf{J} = \Pi \mathbf{B}=0,
  \quad
  \text{since}
  \quad
  \Pi=0,
\end{equation}
for arbitrary characteristics of the external fields and charged particles.
Accounting for the fact that the lowest energy level with $n=0$ does
not contribute to the anomalous current either (see Eq.~(\ref{eq:Jpn=00003D0})
and Ref.~\cite{Dvo17a}), we get that there is no electric current
$\mathbf{J}\parallel\mathbf{B}$ in the system of massive electrons with
anomalous magnetic moments, electroweakly interacting with background
matter, in the state of equilibrium.

Let us comment on the cancellation of the contribution of the lowest energy level to the current $\mathbf{J}\parallel\mathbf{B}$ in Eq.~\eqref{eq:Jpn=00003D0}, where we use the equilibrium Fermi-Dirac distribution function $f_\mathrm{eq}(E) = \left[\exp(\beta E)+1\right]^{-1}$ to eliminate the ultraviolet divergence of the integral. Another regularization scheme was applied in Refs.~\cite{FukKhaWar08,BubGubZhu17}. This scheme is based on imposing the momentum cut-off: $|p_z|<P_\mathrm{max}$, where $P_\mathrm{max} \gg \max(m,\mu B,V_5)$, and setting $f\to1$. This regularization is equivalent to the replacement of the equilibrium distribution function $f_\mathrm{eq}$ by the nonequilibrium one,
\begin{equation}\label{eq:nonequil}
  f_\text{non-eq}(p_z) =
  \begin{cases}
    1, \quad \text{if} \quad |p_z|<P_\mathrm{max},\\
    0, \quad \text{if} \quad |p_z|>P_\mathrm{max}.
  \end{cases}
\end{equation}
The computation of the integral in Eq.~\eqref{eq:Jpn=00003D0} with $f_\text{non-eq}(p_z)$ in Eq.~\eqref{eq:nonequil} gives $\mathbf{J}_{e}^{(n=0)} = - e^{2} V_5 \mathbf{B}/2\pi^{2}$, which formally coincides with the prediction of the CME if we replace $\mu_5 \to -V_5$ in $\mathbf{J}_\mathrm{CME}$. It is interesting to mention, that, in case of $V_5 = 0$, the regularization scheme used in Ref.~\cite{FukKhaWar08} and that applied in Eq.~\eqref{eq:Jpn=00003D0} give coinciding results, $\mathbf{J}_{e}^{(n=0)}=0$, for massive particles.

It is, however, known that, if the distribution function of a system happens to differ from $f_\mathrm{eq}(E)$, the system becomes unstable and will tend to the equilibrium state, i.e. $f_\text{non-eq}(p_z) \to f_\mathrm{eq}(E)$. If we consider an ultrarelativistic plasma, the relaxation time for the process $f_\text{non-eq}(p_z) \to f_\mathrm{eq}(E)$ was estimated in Ref.~\cite{Akh75} as $\tau \sim 10 T^2 /n$, where $n$ is the number density of charged particles in plasma. For instance, in the case of a degenerate electron gas in a neutron star with $T=10^9\,\text{K}$ and $n=10^{36}\,\text{cm}^{-3}$, we get that $\tau\sim10^{-25}\,\text{s}$. Such a small relaxation time is the indication that one should use $f_\mathrm{eq}$ while calculating $\mathbf{J}\parallel\mathbf{B}$ in astrophysical media. Since the typical life-time of a neutron star is much longer than the estimated $\tau$, the gas of massive electrons should be in the equilibrium and, thus, the anomalous current is vanishing, as shown above.

We can consider a situation when the phases with broken and unbroken chiral symmetry coexist. It can happen if the first order chiral phase transition takes place and a bubble with an unbroken phase appears in a neutron star. Then, in the vicinity of the bubble wall, the chiral symmetry can be considered as approximately broken, i.e. $\mathbf{J}_\mathrm{CME} \sim \mu_5 \mathbf{B}$ exists, however, $\mu_5$ is no longer constant. The evolution of $\mu_5$ in this case is driven by the Adler anomaly completed by the chirality flip term~\cite{BoyFroRuc12},
\begin{equation}\label{eq:flip}
  \frac{\mathrm{d}\mu_5}{\mathrm{d}t} =
  \dotsb - \Gamma_f \mu_5,
\end{equation}
where $\Gamma_f $ is the chirality flip rate, which was calculated in Refs.~\cite{GraKapRed15,Dvo16} in the degenerate matter of a neutron star as $\Gamma_f \sim 10^{11}\,\text{s}^{-1} \times (T/10^8\,\text{K})$. In Eq.~\eqref{eq:flip}, we omitted the terms containing the magnetic fields. Taking $T = 10^8\,\text{K}$, one gets that the chiral imbalance relaxation time $\tau = 1/\Gamma_f \sim 10^{-11}\,\text{s}$, which is again much shorter than the neutron star life-time. It means that, even if the chiral symmetry is considered approximately broken, the contribution of the chiral imbalance to the CME is negligible in astrophysical media.

%It is the indication that one cannot expect the enhancement of astrophysical magnetic fields driven by the CME in the system of massive particles.

The opposite situation is implemented if we discuss a quark plasma formed in a collision of heavy ions. In this case, we can take $T\sim10^2\,\text{MeV}$ and $n\sim10^{38}\,\text{cm}^{-3}$. The relaxation time for the distribution function $f_\mathrm{eq}\to f_\mathrm{non-eq}$ is $\tau\sim10^{-12}\,\text{cm}$, which is much greater than a nucleus radius $r_\mathrm{N} \sim 10^{-13}\,\text{cm}$. It means that a plasma emerging in such a collision is strongly nonequilibrium and there is a chance to generate $\mathbf{J}=\Pi\mathbf{B}$ with $\Pi\neq0$.
%even without demanding the chiral symmetry restoration.
This fact is in agreement with the possibility of the appearance of strong magnetic fields in heavy ion collisions~\cite{Kha15}.

%Therefore, if one considers the possibility of the magnetic field generation in astrophysical or cosmological media, which are in the equilibrium state, driven by the current $\mathbf{J}\parallel\mathbf{B}$ of massive particles, one should rely on $f_\text{eq}(E)$. It will give one a vanishing induced current.

The main reason for the disappearance of the current $\mathbf{J}\parallel\mathbf{B}$ in the state of equilibrium 
is the range of the $p_{z}$ variation for massive particles: $-\infty<p_{z}<+\infty$.
This fact distinguishes the considered situation from the CME, where,
for massless particles, one has either $-\infty<p_{z}<0$ or $0<p_{z}<+\infty$,
depending on the particle chirality and its charge.

Now let us compare the obtained result that $\mathbf{J}\parallel\mathbf{B}$ is vanishing with the previous findings. In Ref.~\cite{Dvo17a}, the general expression for the electron current
in Eq.~(\ref{eq:Jtot}) was obtained correctly (see Eq.~(9) in Ref.~\cite{Dvo17a}).
However, while considering the particular case of the degenerate electron gas,
the error in integration over $p_{z}$ was made, that lead to a nonzero $\mathbf{J}\parallel\mathbf{B}$.

It is also interesting to compare our results with those in Ref.~\cite{BubGubZhu17},
where a massive electron with the anomalous magnetic moment, interacting
with an external magnetic field and the axial-vector field $b^{\mu}=(b^{0},\mathbf{b})$,
was considered. The Lagrangian for the electron interaction with the field
$b^{\mu}$ was taken as $\mathcal{L}_{b}=-\bar{\psi}\gamma^{5}\gamma^{\mu}b_{\mu}\psi$.
Comparing $\mathcal{L}_{b}$ with Eq.~(\ref{eq:Larg}), we get that
$b^{0}=V_{5}$. Nonzero spatial components $\mathbf{b}$ can be present
if polarized or moving background matter is considered. %; cf. Ref.~\cite{DvoStu02}.

It was claimed in Ref.~\cite{BubGubZhu17} that, in the situations
(i) $\left(\mu=0,b^{0}\neq0,\mathbf{b}=0\right)$; and (ii) $\left(\mu\neq0,b^{0}\neq0,\mathbf{b}=0\right)$,
there is a nonzero electric current $\mathbf{J}\parallel\mathbf{B}$ in
the system of massive electrons. The case~(i) was previously considered
in our work~\cite{Dvo16a}, where the induced current was shown
to vanish for massive charged particles. Note that neither the lowest
nor higher energy levels contribute to the current. The situation~(ii)
is considered in the present work. Although the cancellation of the
current is not so straightforward as in the case~(i), the current
$\mathbf{J}\parallel\mathbf{B}$ turns out to be vanishing for massive electrons nonetheless.
It should be also noted that any energy level does not contribute
to the current. This finding is again in the contradiction with the
results of Ref.~\cite{BubGubZhu17}.

The reason of the discrepancy of our results with those of Ref.~\cite{BubGubZhu17}
is the following. The anomalous current in Ref.~\cite{BubGubZhu17}
was calculated in vacuum. The regularization was used to eliminate
the ultraviolet divergence in the integrals. We have already mentioned above that this regularization
is equivalent to the consideration of a nonequilibrium state of the system described by the distribution function in Eq.~\eqref{eq:nonequil}. If one makes electrons, forming this current, to thermalize, then $f_\text{non-eq} \to f_\text{eq}$ very rapidly.\footnote{Unless one considers the situation analogous to a collision of heavy ions.} Calculations in the present work
are free of the ultraviolet divergencies since we consider the
system of electrons in the thermodynamic equilibrium with the nonzero temperature $T$ and the chemical
potential $\chi$. The distribution function, used in Eq.~(\ref{eq:Jtot}),
eliminates the appearance of the ultraviolet divergencies. In fact,
this distribution function serves as a natural regularization.

\section{Conclusion\label{sec:CONCL}}

In this work, we have analyzed the possibility of the existence of
the electric current induced along the external magnetic field in
the system of massive charged electrons, having anomalous magnetic
moments and electroweakly interacting with background matter, which
was supposed to be nonmoving and unpolarized. Using the exact solution
of the Dirac equation in the corresponding external fields, which
was obtained in Refs.~\cite{BalStuTok12,BalStuTok13} (see also Appendix~\ref{sec:POSCURR}),
the most general expression of the current, which accounts for the
contributions of both electrons and positrons, has been derived.

The energy of an electron in the external fields in question is quantized
(see Eqs.~(\ref{eq:En>0}) and~(\ref{eq:En=00003D0})) and depends
on the discrete quantum number $n=0,1,\dotsc$. The cancellation of
the contribution of the lowest energy level with $n=0$ to the induced
current can be established directly from the Dirac equation (see Eqs.~(\ref{eq:C13})
and~(\ref{eq:Jpn=00003D0})) even without analyzing the spin integral
of the Dirac equation in Eq.~(\ref{eq:spinoper}). It is important to suppose that the system is in the equilibrium state while considering the cancellation of $\mathbf{J}^{(n=0)}$ (see the discussion in Sec.~\ref{sec:CANCCUR}).

The analysis of the contribution of the higher energy levels with
$n>0$ to the induced current is not trivial. Nevertheless, in Sec.~\ref{sec:CANCCUR},
we have revealed that this contribution is vanishing; cf. Eq.~(\ref{eq:canccur}).
This result is valid at any characteristics of the electron-positron
field, such as $m$, $\mu$, etc., and any parameters of the external
fields, like $B$ and $V_{5}$.

The cancellation of the induced current $\mathbf{J}\parallel\mathbf{B}$
for $n>0$ in the considered system, which was supposed to be in the equilibrium,
corrects the recent claims in Refs.~\cite{Dvo17a,BubGubZhu17} that
such a current can be nonzero. The incorrect nonzero expression for
the current, obtained in Ref.~\cite{Dvo17a}, was because of the error
in the integration over the longitudinal momentum in the case of the
degenerate electron gas.
\footnote{Note that the general expression for the current, derived in Ref.~\cite{Dvo17a},
turns out to be correct; cf. Eq.~(\ref{eq:Jpn>0}) and Eq.~(9) in
Ref.~\cite{Dvo17a}.}
The discrepancy between our results and the findings of Ref.~\cite{BubGubZhu17} can be explained
by the consideration of a nonequilibrium state of the system in Ref.~\cite{BubGubZhu17}. Therefore the current $\mathbf{J} \parallel \mathbf{B} \neq 0$, derived in Ref.~\cite{BubGubZhu17}, will tend to zero very rapidly in a realistic medium. The typical relaxation time for the current to vanish in the astrophysical medium was estimated in Sec.~\ref{sec:CANCCUR}.

%Moreover, one of the results,
%obtained in Refs.~\cite{BubGubZhu17}, namely the existence of the
%nonzero current $\mathbf{J}\parallel\mathbf{B}$ for $b^{0}\neq0$, $\mu=0$,
%and $m\neq0$, contradicts our finding in Ref.~\cite{Dvo16a}, where
%this current was shown to be vanishing.

The main reason for the cancellation of the current consists in the
fact that the longitudinal momentum $p_{z}$ can vary from $-\infty$
to $+\infty$ for a particle with a nonzero mass. Even the feature
of the energy spectrum for $n>0$ that it is not symmetric with respect
to the transformation $p_{z}\to-p_{z}$ (see Eq.~(\ref{eq:ER2})),
which Ref.~\cite{Dvo17a} appealed to in order to justify the existence
of the nonzero induced current, does not help to generate $\mathbf{J}\parallel\mathbf{B}\neq0$.
Thus the cases of $m\neq0$ and $m=0$ are different generically.
In the latter situation, the induced current can exist owing to the
CME (see Sec.~\ref{sec:INTR}), which is based on the asymmetric
motion of charged massless particles at the lowest energy level with
respect to the external magnetic field. The difference between the
systems of massive and massless particles consists in the chiral symmetry: it is broken in the former case and restored
in the latter one.

Therefore, one can expect the existence of the induced current $\mathbf{J}\parallel\mathbf{B}$,
and, thus, the instability of the magnetic field in a system in equilibrium
only if fermions, present in this system, have zero masses. Of course,
in this situation, one should somehow restore the chiral symmetry.
This task is not trivial as explained in Sec.~\ref{sec:INTR}.

The nonsmooth behavior of the current $\mathbf{J}\parallel\mathbf{B}$ as a function of the particle mass can be treated as follows. In some circumstances, the induced current itself $\mathbf{J}\parallel\mathbf{B}$ is not the object of a study. For instance, if one considers the generation of magnetic fields, one is interested in the evolution of the magnetic field driven by the current $\mathbf{J}\parallel\mathbf{B}$.
For instance, let us discuss the electron-positron plasma with a nonzero temperature. In this case, the mass of a particle becomes the function of the temperature $m=m(T)$. 
Suppose that a first order chiral phase transition can happen in this plasma at a certain temperature $T_c$. If $T<T_c$, particles are massive and there is no CME in the system.

Let the temperature to increase. As soon as $T=T_c$, bubbles of the new phase with restored chiral symmetry appear in plasma. Since masses of fermions are equal to zero inside a bubble, $\mathbf{J}_\mathrm{CME} \parallel \mathbf{B}$ can flow there, causing the magnetic field instability, which, in its turn, leads to the enhancement of a seed field. Just after the phase transition, the size of these bubbles is small and, hence, the magnetic field is small-scale. However, in the course of time, the scale of the magnetic field will smoothly increase together with the size of bubbles. This qualitative analysis shows that a smooth change of an external parameter, like the plasma temperature, will result in the smooth variation of a measurable quantity, such as the magnetic field length scale, which is related to the CME. In this consideration, the particle mass is an auxiliary parameter.

Note that the disappearance of the current $\mathbf{J}\parallel\mathbf{B}$ of massive fermions,
participating in parity violating interactions,
was mentioned for the first time in Ref.~\cite{Vil80}. That result was obtained perturbatively by considering the loop contribution to the photon polarization tensor.
% in case when charged particles participate in the parity violating interaction.
In the present work we have generalized the finding of Ref.~\cite{Vil80} to a more complex physics system, which also accounts for anomalous magnetic moments of massive charged fermions. Moreover, our approach to demonstrate that $\mathbf{J}\parallel\mathbf{B} = 0$ is nonperturbative since it is based on the exact solution of the Dirac equation in the presence of the external fields.

\section*{Acknowledgments}

%I am thankful to S.B.~Popov, A.I.~Rez, and R.~Turolla for useful communications, to V.B.~Se\-mi\-koz and D.D.Sokoloff for helpful discussions, as well as to
I am thankful to V.~B.~Semikoz for useful comments, to the Tomsk State University Competitiveness Improvement
Program, RFBR (research project No.~18-02-00149a), and the Foundation for the Advancement of Theoretical Physics and Mathematics ``BASIS'' for a partial support.

\appendix

\section{Calculation of the induced current\label{sec:POSCURR}}

In this appendix, we compute the electric currents of both electrons
and positrons. The charged particles are supposed to be massive, have
anomalous magnetic moments, and electroweakly interact with background
matter under the influence of the constant and homogeneous magnetic field.
The computation of the currents is based on the exact solution of
the Dirac equation in the external fields. Earlier this equation for
an electron was solved in Refs.~\cite{BalStuTok12,BalStuTok13} using
the standard representation of the Dirac matrices.

Let us suppose that an electron interacts with the unpolarized and
nonmoving background matter consisting of neutrons and protons and
with the magnetic field, which is along the $z$-axis: $\mathbf{B}=B\mathbf{e}_{z}$.
Then, the Dirac equation, which can be obtained from the Lagrangian
in Eq.~(\ref{eq:Larg}), for an electron wave function $\psi_{e}$
has the form,
\begin{equation}\label{eq:Direq}
  \mathrm{i}\dot{\psi}_{e}=\hat{H}\psi_{e},  
  \quad
  \hat{H}=\left(\bm{\alpha}\mathbf{P}\right)+\beta m +
  \mu B\beta\Sigma_{3} +
  V_{\mathrm{L}}P_{\mathrm{L}}+V_{\mathrm{R}}P_{\mathrm{R}},
\end{equation}
where $\mathbf{P}=-\mathrm{i}\nabla-e\mathbf{A}$ is the canonical
momentum operator, $\mathbf{A}=\left(0,Bx,0\right)$ is vector potential
in the Landau gauge, $\bm{\alpha}=\gamma^{0}\bm{\gamma}$, $\beta=\gamma^{0}$
and $\bm{\Sigma}=\gamma^{0}\bm{\gamma}\gamma^{5}$ are the Dirac matrices.
The effective potentials $V_{\mathrm{L,R}}$ of the electroweak interaction
with matter are defined in Sec.~\ref{sec:CANCCUR}.

Let us look for the solution of Eq.~(\ref{eq:Direq}) in the form,
\begin{equation}\label{eq:psiel}
  \psi_{e}=\exp
  \left(
    -\mathrm{i}Et+\mathrm{i}p_{y}y+\mathrm{i}p_{z}z
  \right)
  \psi_{x},
\end{equation}
where $\psi_{x}=\psi(x)$ is the bispinor which depends on $x$ and
$-\infty<p_{y,z}<+\infty$. We shall choose the chiral representation
of the Dirac matrices~\cite{ItzZub80},
\begin{equation}\label{eq:chirrep}
  \gamma^{\mu} =
  \begin{pmatrix}
    0 & -\sigma^{\mu}\\
    -\bar{\sigma}^{\mu} & 0\ 
  \end{pmatrix},
  \quad
  \sigma^{\mu}=(\sigma_{0},-\bm{\sigma}),
  \quad
  \bar{\sigma}^{\mu}=(\sigma_{0},\bm{\sigma}),
\end{equation}
where $\sigma_{0}$ is the unit $2\times2$ matrix and $\bm{\sigma}$
are the Pauli matrices. Using Eq.~(\ref{eq:chirrep}), we can represent
$\psi_{x}$ in the form,
\begin{equation}\label{eq:psix}
  \psi_{x}^{\mathrm{T}} =
  \left(
    C_{1}u_{n-1},\mathrm{i}C_{2}u_{n},C_{3}u_{n-1},\mathrm{i}C_{4}u_{n}
  \right),
\end{equation}
where $C_{i}$, $i=1,\dots,4,$ are the spin coefficients,
\begin{equation}\label{eq:Hermfun}
  u_{n}(\eta)=
  \left(
    \frac{eB}{\pi}
  \right)^{1/4}
  \exp
  \left(
    -\frac{\eta^{2}}{2}
  \right)
  \frac{H_{n}(\eta)}{\sqrt{2^{n}n!}},
  \quad
  n=0,1,\dotsc,
\end{equation}
are the Hermite functions, $H_{n}(\eta)$ are the Hermite polynomials,
and $\eta=\sqrt{eB}x+p_{y}/\sqrt{eB}$.

The energy levels for $n>0$ were found in Refs.~\cite{BalStuTok12,BalStuTok13}
as
\begin{equation}\label{eq:En>0}
  E=\bar{V}+\lambda\mathcal{E},
  \quad
  \mathcal{E}=\sqrt{p_{z}^{2}+2eBn+m^{2}+(\mu B)^{2}+V_{5}^{2}-2sm|S|},
\end{equation}
where 
\begin{equation}\label{eq:S2}
  |S| = \frac{1}{m}\sqrt{(p_{z}V_{5}-\mu Bm)^{2}+2eBn
  \left[
    (\mu B)^{2}+V_{5}^{2}
  \right]},
\end{equation}
is the absolute value of the eigenvalue of the spin operator~\cite{BalStuTok12,BalStuTok13},
\begin{equation}\label{eq:spinoper}
  \hat{S}=V_{5}\hat{S}_{l}-\mu B\hat{S}_{t},  
  \quad
  \hat{S}_{l}=\frac{(\bm{\Sigma}\cdot\mathbf{P})}{m},
  \quad
  \hat{S}_{t}=\Sigma_{3}-\frac{\mathrm{i}}{m}(\bm{\gamma}\times\mathbf{P})_{3},
\end{equation}
which commutes with the Hamiltonian $\hat{H}$ in Eq.~(\ref{eq:Direq}).
In Eq.~(\ref{eq:En>0}), the discrete quantum number $s=\pm1$ is
the sign of $S$, $\bar{V}=(V_{\mathrm{L}}+V_{\mathrm{R}})/2$, $V_{5}=(V_{\mathrm{L}}-V_{\mathrm{R}})/2$,
and $\lambda=\pm1$ is the sign of the energy, i.e. the electron energy
reads $E_{e}=E(\lambda=+1)=\mathcal{E}+\bar{V}$, and the positron
energy has the form, $E_{\bar{e}}=-E(\lambda=-1)=\mathcal{E}-\bar{V}$.

The spin coefficients for $n>0$ also were found in Refs.~\cite{BalStuTok12,BalStuTok13},
using the standard representation of the Dirac matrices. Since we
choose the chiral representation, which is more convenient for our
purposes, we just briefly list the main results. The spin coefficients
can be represented in the form, 
\begin{align}\label{eq:CiAB}
  \left(
    \begin{array}{c}
      C_{1}\\
      C_{3}
    \end{array}
  \right) & =
  \frac{1}{\sqrt{2}}
  \left(
    1+\frac{p_{z}V_{5}-\mu Bm}{mS}
  \right)^{1/2}
  \left(
    \begin{array}{cc}
      Z & -\mu B/Z\\
      \mu B/Z & Z
    \end{array}
  \right)
  \left(
    \begin{array}{c}
      A_{1}\\
      A_{2}
    \end{array}
  \right),
  \nonumber
  \\
  \left(
    \begin{array}{c}
      C_{2}\\
      C_{4}
    \end{array}
  \right) & =
  \frac{s}{\sqrt{2}}
  \left(
    1-\frac{p_{z}V_{5}-\mu Bm}{mS}
  \right)^{1/2}
  \left(
    \begin{array}{cc}
      Z & \mu B/Z\\
      -\mu B/Z & Z
    \end{array}
  \right)
  \left(
    \begin{array}{c}
      A_{1}\\
      A_{2}
    \end{array}
  \right).
\end{align}
where $Z=\left(V_{5}+\sqrt{V_{5}^{2}+(\mu B)^{2}}\right)^{1/2}$ .
The new auxiliary coefficients $A_{i}$, $i=1,2$, are completely
defined by the following expressions:
\begin{gather}
  A_{1}^{2} =
  \left(
    1+\frac{mS-V_{5}^{2}-(\mu B)^{2}}{\mathcal{E}\sqrt{V_{5}^{2}+(\mu B)^{2}}}
  \right)
  C^{2},
  \quad
  A_{2}^{2}=
  \left(
    1-\frac{mS-V_{5}^{2}-(\mu B)^{2}}{\mathcal{E}\sqrt{V_{5}^{2}+(\mu B)^{2}}}
  \right)
  C^{2},
  \nonumber
  \\
  A_{1}A_{2} =
  - \frac{\mu Bp_{z}+mV_{5}}{\mathcal{E}\sqrt{V_{5}^{2}+(\mu B)^{2}}}C^{2},
  \label{eq:AiC}
\end{gather}
where $C$ is the normalization coefficient having the form,
\begin{equation}\label{eq:C2}
  C^{2}=\frac{1}{4(2\pi)^{2}\sqrt{V_{5}^{2}+(\mu B)^{2}}},
\end{equation}
which can be found if we normalize the total wave function as
\begin{equation}\label{eq:psinorm}
  \int\mathrm{d}^{3}x\psi_{p_{y}p_{z}n}^{\dagger}\psi_{p'_{y}p'_{z}n'} =
  \delta\left(p_{y}-p'_{y}\right)\delta\left(p_{z}-p'_{z}\right)\delta_{nn'}.
\end{equation}
We can see that Eqs.~(\ref{eq:CiAB})-(\ref{eq:C2})
completely define the spin coefficients $C_{i}$ at $n>0$.

If $n=0$, we should set $C_{1}=C_{3}=0$ to avoid the appearance
of the Hermite functions with negative indexes in Eq.~(\ref{eq:psix}).
The energy spectrum reads
\begin{equation}\label{eq:En=00003D0}
  E=\bar{V}+\lambda\mathcal{E},
  \quad
  \mathcal{E}=\sqrt{(p_{z}+V_{5})^{2}+(m-\mu B)^{2}}.
\end{equation}
The nonzero spin coefficients $C_{2,4}$ have the form,
\begin{equation}\label{eq:C13}
  C_{2}^{2} =
  \frac{1}{2\mathcal{E}(2\pi)^{2}}\frac{(m-\mu B)^{2}}{\mathcal{E}+p_{z}+V_{5}},
  \quad
  C_{4}^{2}=\frac{\mathcal{E}+p_{z}+V_{5}}{2\mathcal{E}(2\pi)^{2}},
\end{equation}
where we use the normalization condition in Eq.~(\ref{eq:psinorm}).

The electric current of electrons has the form,
\begin{equation}\label{eq:Jzgen}
  \mathbf{J}_{e} = -e
  \sum_{n=0}^{\infty}\sum_{s}
  \int_{-\infty}^{+\infty}\mathrm{d}p_{y}\mathrm{d}p_{z}
  \bar{\psi}_{e}\bm{\gamma}\psi_{e}
  f(E_{e}-\chi),
\end{equation}
where $f(E)=\left[\exp(\beta E)+1\right]^{-1}$ is the Fermi-Dirac
distribution function, $\chi$ is the chemical potential, and $\beta=1/T$
is the reciprocal temperature. If the electron wave functions are treated as secondly quantized objects, we should define the current in Eq.~\eqref{eq:Jzgen} using the normal ordering $\colon\mathbf{J}_{e}\colon$ to remove the infinite contribution of the vacuum energy. The normal ordering of creation and annihilation operators should be performed accounting for the external fields, i.e. Fock states are defined in the presence of the magnetic field and the electroweak interaction with background matter~\cite{FraGitShv91}.

First, we notice that
\begin{equation}\label{eq:dx}
  \int_{-\infty}^{+\infty}\mathrm{d}p_{y}\bar{\psi}_{e}\bm{\gamma}\psi_{e} =
  eB\int_{-\infty}^{+\infty}\mathrm{d}x\psi_{e}^{\dagger}\bm{\alpha}\psi_{e}.
\end{equation}
Thus $J_{x,y}\sim\left\langle \psi_{e}^{\dagger}\alpha_{1,2}\psi_{e}\right\rangle =0$
since $\alpha_{1,2}=\text{diag}(\sigma_{1,2},-\sigma_{1,2})$ and
$\sigma_{1,2}$ are the nondiagonal Pauli matrices. Indeed, the integration
over $x$ in Eq.~(\ref{eq:dx}) is vanishing for $J_{x,y}$ owing to the orthogonality
of Hermite functions with different indexes,
\begin{equation}\label{eq:unorm}
  \int_{-\infty}^{+\infty}\mathrm{d}xu_{n}(\eta)u_{n'}(\eta)=\delta_{nn'}.
\end{equation}
Therefore only the component of the current along the magnetic field
$J_{z}$ should be analyzed.

Then, let us consider the contribution to $J_{z}$ from the higher
energy levels with $n>0$. In the chiral representation of the Dirac
matrices, with help of Eqs.~(\ref{eq:dx}) and~(\ref{eq:unorm}),
we have
\begin{equation}\label{eq:dpy}
  \int_{-\infty}^{+\infty}\mathrm{d}p_{y}\bar{\psi}_{e}\gamma^{3}\psi_{e} =
  eB
  \left(
    C_{1}^{2}+C_{4}^{2}-C_{2}^{2}-C_{3}^{2}
  \right).
\end{equation}
Using Eqs.~(\ref{eq:CiAB})-(\ref{eq:C2}),
we obtain that
\begin{align}\label{eq:Cicomb}
  C_{1}^{2}+C_{4}^{2}-C_{2}^{2}-C_{3}^{2}= &
  -4\mu BA_{1}A_{2}+2V_{5}\frac{\mu Bm-V_{5}p_{z}}{m\tilde{S}}
  \left(
    A_{1}^{2}-A_{2}^{2}
  \right)
  \nonumber
  \\
  & =
  \frac{1}{(2\pi)^{2}\mathcal{E}}
  \left[
    p_{z}
    \left(
      1-\frac{V_{5}^{2}}{mS}
    \right) +
    V_{5}\frac{\mu B}{S}
  \right].
\end{align}
Introducing the quantity $R^{2}=m|S|$, changing $s\to-s$, and returning
to the vector notations, we get the contribution of the higher energy
levels to the current as
\begin{equation}\label{eq:Jpn>0}
  \mathbf{J}_{e}^{(n>0)} =
  -\frac{e^{2}\mathbf{B}}{(2\pi)^{2}}
  \sum_{n=1}^{\infty}
  \sum_{s=\pm1}\int_{-\infty}^{+\infty}
  \frac{\mathrm{d}p_{z}}{\mathcal{E}}
  \left[
    p_{z}
    \left(
      1+s\frac{V_{5}^{2}}{R^{2}}
    \right) -
    s\frac{\mu BmV_{5}}{R^{2}}
  \right]
  f(\mathcal{E}-\chi_{\mathrm{eff}}),
\end{equation}
where $\chi_{\mathrm{eff}}=\chi-\bar{V}$. Eq.~\eqref{eq:Jpn>0} is used in Eq.~\eqref{eq:Jtot}.

The contribution of the lowest energy level with $n=0$ to the current
can be obtained on the basis of Eqs.~(\ref{eq:En=00003D0}), (\ref{eq:C13}),
and~(\ref{eq:dpy}) as
\begin{align}\label{eq:Jpn=00003D0}
  \mathbf{J}_{e}^{(n=0)}= & e^{2}\mathbf{B}
  \int_{-\infty}^{+\infty}\mathrm{d}p_{z}f(E_{e}-\chi)
  \left(
    C_{2}^{2}-C_{4}^{2}
  \right)
  \nonumber
  \\
  & =
  -\frac{e^{2}\mathbf{B}}{(2\pi)^{2}}
  \int_{-\infty}^{+\infty}\mathrm{d}p_{z}
  f(E_{e}-\chi)\frac{p_{z}+V_{5}}{\sqrt{(p_{z}+V_{5})^{2}+(m-\mu B)^{2}}}=0,
\end{align}
since the integrand is the odd function. One can see in Eq.~(\ref{eq:Jpn=00003D0})
that the contribution to the electric current from the lowest energy
level vanishes. This result is valid for any characteristics of the
external fields and charged particles. This our finding extends the
result of Ref.~\cite{Dvo16a} to the situation when the anomalous
magnetic moment of charged particles is taken into account.

The positron wave function can be obtained on the basis of Eq.~(\ref{eq:psiel})
by applying the charge conjugation operation, $\psi_{\bar{e}}=\mathrm{i}\gamma^{2}\psi_{e}^{*}$,
and setting $\lambda=-1$ in the energy spectrum. Finally, using Eqs.~(\ref{eq:psiel})
and~(\ref{eq:psix}) we get
\begin{align}\label{eq:psipos}
  \psi_{\bar{e}}^{\mathrm{T}}= & 
  \exp
  \left(
    -\mathrm{i}E_{\bar{e}}t-\mathrm{i}p_{y}y-\mathrm{i}p_{z}z
  \right)
  \nonumber
  \\
  & \times
  \left(
    -\mathrm{i}C_{4}u_{n},-C_{3}u_{n-1},\mathrm{i}C_{2}u_{n},C_{1}u_{n-1}
  \right),
\end{align}
where the coefficients $C_{i}$ are defined by Eqs.~(\ref{eq:CiAB})-(\ref{eq:C2}).

The expression for the current of positrons has the form, 
\begin{equation}\label{eq:Jposgen}
  \mathbf{J}_{\bar{e}}=e
  \sum_{n=0}^{\infty}\sum_{s}\int_{-\infty}^{+\infty}
  \mathrm{d}p_{y}\mathrm{d}p_{z}\bar{\psi}_{\bar{e}}\bm{\gamma}\psi_{\bar{e}}
  f(E_{\bar{e}}+\chi).
\end{equation}
Analogously to the electron case, we obtain that the transversal (with
respect to $\mathbf{B}$) components of $\mathbf{J}_{\bar{e}}$ are
equal to zero. Using Eqs.~(\ref{eq:dpy}), (\ref{eq:Cicomb}), and~(\ref{eq:psipos}),
we get on the basis of Eq.~(\ref{eq:Jposgen}) that
\begin{equation}\label{eq:Jposn>0}
  \mathbf{J}_{\bar{e}}^{(n>0)} =
  \frac{e^{2}\mathbf{B}}{(2\pi)^{2}}
  \sum_{n=1}^{\infty}\sum_{s=\pm1}\int_{-\infty}^{+\infty}
  \frac{\mathrm{d}p_{z}}{\mathcal{E}}
  \left[
    p_{z}
    \left(
      1+s\frac{V_{5}^{2}}{R^{2}}
    \right) -
    s\frac{\mu BmV_{5}}{R^{2}}
  \right]
  f(\mathcal{E}+\chi_{\mathrm{eff}}).
\end{equation}
Comparing Eqs.~(\ref{eq:Jpn>0}) and~(\ref{eq:Jposn>0}), we can
see that the positrons current flows in the opposite direction and
has the opposite sign of $\chi_{\mathrm{eff}}$ in the distribution
function. Eq.~\eqref{eq:Jposn>0} is used in Eq.~\eqref{eq:Jtot}.

The contribution of the lowest energy level with $n=0$ to $\mathbf{J}_{\bar{e}}$
can be obtained analogously to $\mathbf{J}_{e}$ by setting $C_{1}=C_{3}=0$
in Eq.~(\ref{eq:psipos}). One obtains that $\mathbf{J}_{\bar{e}}^{(n=0)}=-\mathbf{J}_{e}^{(n=0)}=0$;
cf. Eq.~(\ref{eq:Jpn=00003D0}).

There is a special case when $m=\mu B$. In this situation, particles at the lowest energy level with $n=0$ become effectively massless; cf. Eq.~\eqref{eq:En=00003D0}. The wave function of a ``left" electron, which satisfies the normalization condition in Eq.~\eqref{eq:psinorm}, reads
\begin{equation}\label{eq:psieL}
  \psi_{e\mathrm{L}}^\mathrm{T}=
  \frac{\mathrm{i}u_0}{2\pi}
  \exp
  \left(
    -\mathrm{i}E_{e\mathrm{L}}t+\mathrm{i}p_{y}y+\mathrm{i}p_{z}z
  \right)
  \times
  (0,0,0,1),
\end{equation}
where the energy level has the form,
\begin{equation}\label{eq:EeL}
  E_{e\mathrm{L}} = p_z + V_\mathrm{L},
  \quad
  - V_5 < p_z < + \infty,
\end{equation}
Analogously for ``right" electrons one has,
\begin{equation}\label{eq:psieR}
  \psi_{e\mathrm{R}}^\mathrm{T}=
  \frac{\mathrm{i}u_0}{2\pi}
  \exp
  \left(
    -\mathrm{i}E_{e\mathrm{R}}t+\mathrm{i}p_{y}y+\mathrm{i}p_{z}z
  \right)
  \times
  (0,1,0,0),
\end{equation}
and
\begin{equation}\label{eq:EeR}
  E_{e\mathrm{R}} = - p_z + V_\mathrm{R},
  \quad
  - \infty < p_z < - V_5.
\end{equation}
One can see in Eqs.~\eqref{eq:EeL} and~\eqref{eq:EeR} that the range of the $p_z$ variation becomes not symmetric like in the case of the CME~\cite{Vil80,NieNin81,FukKhaWar08}.

The electric current of electrons can be computed using Eqs.~\eqref{eq:Jzgen}-\eqref{eq:unorm} and~\eqref{eq:psieL}-\eqref{eq:EeR} as
\begin{equation}\label{eq:Jespeccase}
  \mathbf{J}_{e}= \frac{e^{2}\mathbf{B}}{(2\pi)^{2}}
  \left[
    \int_{-V_5}^{+\infty} \mathrm{d} p_z f(p_z + V_\mathrm{L} - \chi) % left contribution
    - 
    \int_{-\infty}^{-V_5} \mathrm{d} p_z f(- p_z + V_\mathrm{R} - \chi) % right contribution
  \right].
%  \notag
%  \\
%  & =
%  \int_{0}^{+\infty}
%  \left[
%    f(p + \bar{V} - \chi_\mathrm{L}) - f(p + \bar{V} - \chi_\mathrm{R})
%  \right]
\end{equation}
Changing the variables $p_z \to p = p_z - V_5$ and $p_z \to p = - p_z - V_5$ in the first and the second integrals in Eq.~\eqref{eq:Jespeccase} respectively, we can see that $\mathbf{J}_{e}=0$. Analogously we can show that $\mathbf{J}_{\bar{e}} = 0$. Thus, even in this special situation, when $B=m/\mu$, the induced current $\mathbf{J}\parallel\mathbf{B}$ is absent.

\end{document}